\documentclass[aip,jcp,twocolumn,superscriptaddress,groupedaddress]{revtex4}  
\usepackage{amsfonts}
\usepackage{amsmath}
\usepackage{amssymb}
\usepackage{version}
\usepackage{graphicx}%
\includeversion{New_connection}
\excludeversion{Old_connection}

\begin{document}
\title{Constant pressure and temperature discrete-time Langevin molecular dynamics}
\author{Niels Gr{\o}nbech-Jensen}
\affiliation{Department of Mechanical and Aerospace Engineering, University of California, Davis, CA 95616}
\affiliation{Department of Mathematics, University of California, Davis, CA 95616}
\author{Oded Farago}
\affiliation{Department of Biomedical Engineering, Ben Gurion University of the Negev, Be'er Sheva, 84105 Israel}
\affiliation{Ilse Katz Institute for Nanoscale Science and Technology, Ben Gurion University of the Negev, Be'er Sheva, 84105 Israel}
\begin{abstract}
We present a new and improved method for simultaneous control of
temperature and pressure in molecular dynamics simulations with
periodic boundary conditions. The thermostat-barostat equations are
build on our previously developed stochastic thermostat, which has
been shown to provide correct statistical configurational sampling for
any time step that yields stable trajectories. Here, we extend the
method and develop a set of discrete-time equations of motion for both
particle dynamics and system volume in order to seek pressure control
that is insensitive to the choice of the numerical time step.  The
resulting method is simple, practical, and efficient. The method is
demonstrated through direct numerical simulations of two
characteristic model systems - a one dimensional particle chain for
which exact statistical results can be obtained and used as
benchmarks, and a three dimensional system of Lennard-Jones
interacting particles simulated in both solid and liquid phases. The
results, which are compared against the method of Kolb \& D\"{u}nweg,
show that the new method behaves according to the objective, namely
that acquired statistical averages and fluctuations of configurational
measures are accurate and robust against the chosen time step applied
to the simulation.
\end{abstract}
\maketitle

\section{Introduction}
\label{sec:intro}

Molecular Dynamics (MD) computer simulations have become a standard
tool for investigating a variety of atomic and molecular systems
ranging from solids to simple fluids to complex biomolecular
assemblies \cite{rapaport}. They are particularly attractive for
dynamics and for equilibrium sampling in high density condensed matter
systems where large scale collective modes may be significant. These
modes may not be easily excited (and relaxed) by the alternative
approach to phase space sampling, namely Monte Carlo (MC) simulations,
because: (i) MC evolution is diffusive in nature, and (ii) MC tends to
have low acceptance rates in high density regions \cite{binder}. The
MC method, however, possesses one significant advantage over MD - the
ability to sample, at least in principle (i.e., for sufficiently long
runs), almost any statistical ensemble in a fairly straightforward
manner. This task is accomplished by performing MC moves that are
ergodic and satisfy the detailed balance condition. In the canonical
$(N,V,T)$ ensemble (where $N$, $V$, and $T$ denote the number of
particles, volume, and temperature, respectively), the latter
requirement is usually fulfilled by using the Metropolis criterion for
the acceptance probability: $p_{\rm acc}=\min [1,\exp(-\Delta
U/k_BT)]$, where $k_B$ is the Boltzmann constant and $\Delta U$ is the
change in the potential energy between the two states which are
approached via opposite moves. Similarly, the isothermal-isobaric
$(N,P,T)$ (where $P$ is the pressure) ensemble can be simulated by
including coordinate displacements that change the volume of the
system and scale the coordinates of the particles accordingly, and
by redefining the potential energy to $U_{\rm eff}=U+PV-Nk_BT\ln{V}$
[see, later, Eq.~(\ref{eq:ueff})].

Things become more complicated when it comes to MD simulations, which
attempt to follow the {\it dynamics} of a molecular system by
numerically solving Newton's classical equations of motion for the
constituent particles. This, supposedly, generates trajectories
within the microcanonical ensemble $(N,V,E)$ (where $E$ is the
internal energy of the system, which is the sum of potential and
kinetic energies) although, due to truncation errors, one should not
expect the computed trajectories to actually follow the real-time
dynamics in many-particle systems \cite{frenkel_smit}. The most
commonly used discrete-time integrator for MD simulations is the
St\"{o}rmer-Verlet algorithm
which (in its so-called ``velocity-Verlet'' form) reads \cite{verlet}:
\begin{eqnarray}
r^{n+1}&=&r^n+v^ndt+\frac{dt^2}{2m}f^n \label{eq:verlet_r}
\\
v^{n+1}&=&v^n+\frac{dt}{2m}\left(f^n+f^{n+1}\right),
\label{eq:verlet_v}
\end{eqnarray}
where $r^n$, $v^n$, and $f^n$ denote, respectively, the coordinate,
velocity, and the force acting on the particle with mass $m$ at time
$t_n$, and $t_{n+1}=t_n+dt$. Notice that $r^n$, $v^n$, and $f^n$
represent Cartesian components, which means that for a system of $N$
particles in a space with dimensionality $d$, the number of equations
one needs to compute per time step $dt$, is $2dN$.  The Verlet
algorithm results in a trajectory which is accurate to second order in
the time step $dt$. This deviation between the computed and the
``correct'' trajectories should not be a matter of concern if the
simulations properly sample the correct statistical ensemble, or
otherwise retain the measures of interest. Thus, the critical test for
the performance of any numerical integrator must be its accuracy in
measuring important thermodynamic quantities and the variations of the
results with $dt$. Remarkably, the Verlet algorithm suffers from the
problem that the total kinetic energy of a simulated system (which is
supposed to be proportional to the temperature) becomes progressively
depressed for increasing time step $dt$ \cite{eastwood,gjf} compared
to the potential energy (see also Appendix \ref{appendix_b} for a
harmonic oscillator demonstration of how the velocity in discrete time
is not precisely the velocity of the corresponding spatial
coordinate). Other thermodynamic observables also exhibit variations
with $dt$, which makes a striking contrast with MC simulations in
which the thermodynamic configurational averages are insensitive to
the step sizes.

The microcanonical ensemble sampled in MD simulations does not provide
the best representation of experimental conditions, where the most
common condition is that of constant temperature and
pressure. Therefore, considerable effort has been devoted to the
development of MD algorithms for simulations of the
isothermal-isobaric ensemble. In the simplest method, proposed by
Berendsen, the system is weakly coupled to external heat
(``thermostat'') and pressure (``barostat'') baths, using the
principle of least local perturbation \cite{berendsen}. This method
has been criticized for failing to correctly sample the statistical
ensemble, due to its tendency to suppress fluctuations in kinetic
energy and volume. A second, more reliable method, pioneered by
Andersen for fixed pressure \cite{andersen}, extended by Parrinello and
Rahman \cite{parrinello} and by Nos\'{e}
\cite{nose}, and revised by Hoover \cite{hoover} to fixed temperature
MD simulations, is the extended Lagrangian formalism. The method is
based on the idea of including additional degrees of freedom,
corresponding to the volume and/or the kinetic energy of the system,
together with their conjugate momenta variables. The new variables are
coupled to the system in a manner which guarantees that the trajectory
correctly samples the isothermal-isobaric ensemble. The latter
constitutes a sub-space of the configuration space of the extended
system.  Within the extended phase space, the statistics is
microcanonical and the equations of motion can be derived from the
extended Hamiltonian, which is conserved in time. In principle, these
Hamiltonian equations of motion can be integrated numerically using
the Verlet algorithm. In practice, the implementation of the
discrete-time Verlet algorithm raises several significant challenges
and difficulties. Specifically, for the barostat part, the coupling
between the particles' degrees of freedom and the ``piston''
(introduced to control the volume fluctuations) is the source of the
following problems:
\begin{enumerate}
\item When the piston moves, the particle coordinates must be
rescaled, which leads to a metric problem with the algorithm. This
problem has been addressed in Refs.~\cite{hoover2,hoover3,klein}.
\item The method is extremely sensitive to the value assigned to the
mass of the piston.  A low mass will result in rapid box size
oscillations which are not attenuated very efficiently by the motions of
the molecules, while a large mass will give rise to a slow adjustment
of the volume and may therefore be computationally inefficient.
\item The force on the piston [See Eq.~(\ref{eq:presandersen}) below]
depends on the internal pressure of the system, the value of which
depends on the instantaneous kinetic energy of the particles. This
means that the velocity $v^{n+1}$ is the solution to an implicit
equation, which therefore must be solved iteratively. This has several
consequences, including that the computed trajectory is no longer time
reversible - a feature that jeopardizes the (extended) energy
conservation in long simulations. A set of explicit reversible
integrators for the dynamics has been developed by Martyna and
coworkers \cite{martyna,tuckerman1,tuckerman2}.
\item The dependence of the internal pressure on the kinetic energy
leads to inaccurate determination of the pressure, since the kinetic
contribution is derived from the particles' velocities, which, as
shown in Appendix \ref{appendix_b}, deviate from the actual
velocities.
\end{enumerate}

The third approach to constant pressure and temperature MD simulations
employs the Andersen extended Lagrangian formalism, i.e., it
couples the system to a global piston which
governs the volume fluctuations of the system. However, instead of
using a Nos\'{e}-Hoover thermostat and solving the Hamiltonian equations
of motion, the temperature is set by solving the Langevin equation
\cite{coffey}:
\begin{eqnarray}
\dot{r}&=&v \\
m\dot{v}&=&f(r,t)-\alpha v+\beta(t).
\end{eqnarray}
The Langevin equation describes Newtonian dynamics where the conservative
force field $f(r,t)$ is augmented by: (i) a friction force
proportional to the velocity with friction coefficient $\alpha$, and
(ii) thermal white (``delta-function correlated'') noise,
$\beta(t)$. The friction and noise terms represent the interactions
with the implicit degrees of freedom of the heat bath. In order to
satisfy Einstein's fluctuation-dissipation theorem that relates the
friction and noise to each other, it is usually assumed that the noise
is Gaussian distributed and has the following statistical properties
\cite{parisi}:
\begin{eqnarray}
\langle \beta(t)\rangle&=&0
\label{eq:noise1} \\
\langle\beta(t)\beta(t^{\prime})\rangle&=&2\alpha
k_BT\delta(t-t^{\prime}).
\label{eq:noise2}
\end{eqnarray}
Historically, Langevin stochastic thermostats have been developed
in parallel to the Nos\'{e}-Hoover deterministic thermostat, in the
early 80' \cite{bbk}. However, it was only in 1995, when Feller {\em
et al.}\/~proposed to simulate isothermal-isobaric conditions by
considering Langevin dynamics for the piston's equation of motion in
Andersen's extended system \cite{feller}. This approach was improved a
few years later by Kolb and D\"{u}nweg, who considered Langevin
dynamics for both the particles and piston, and who developed an
integrator for this purpose
\cite{kolb}. While many of the problems associated with the
application of the Nos\'{e}-Hoover thermostat for $(N,P,T)$
simulations remained unsolved (especially those originating from the
coupling between the movement of the piston and the particles), the
idea of simulating the extended system withing the framework of
Langevin dynamics appears to offer shorter correlation times and
improved sampling.

In this paper we present a new method for Langevin dynamics
simulations at constant pressure and temperature. The method, which is
both effective and simple to implement, provides improvements compared
to the method of Kolb and D\"{u}nweg (KD). Key distinctions between
our algorithm and others, including KD, lie in the manner by which the
displacements of the particles into ``physical'' and ``scaled''
components are decoupled. Within the traditional methods these two
displacements are defined and separated prior to time-discretization,
while our method is based on formulating the equations of motion for
an already temporally discretized set of coordinates. Another change
that we introduce in the method, is to replace the kinetic energy term
in the instantaneous pressure with its known thermodynamic average,
which is precisely the ideal gas pressure $Nk_BT/V$. This change does
not only resolve the aforementioned problems in implementations of
Verlet-type integrators to the piston's equation of motion, it also
makes the extended Lagrangian dynamics {\em more}\/ consistent with
the statistical mechanics of the isothermal-isobaric ensemble that the
simulations aims to sample. Finally, we take advantage of the recent
advances in numerical integrators for Langevin dynamics and replace
the old BBK (Brooks, Br\"{u}nger, and Karplus) thermostat \cite{bbk}
with the recently introduced G-JF (Gr{\o}nbech-Jensen and Farago)
thermostat \cite{gjf}. While the former has a simulated temperature
that differs by ${\cal O}(dt)$ from the correct one, the latter
exhibits no detectable changes in the configurational sampling
statistics as the time step is varied in the entire numerical
stability range \cite{gjf,gjf2}.

The paper is organized as follows: In section \ref{sec:method} we
derive the new method for isothermal-isobaric MD simulations. This
section contains both a detailed discussion of the theoretical aspects
of the method, as well as a derivation of the algorithm for
isothermal-isobaric simulations. The new algorithm is tested against
the method of Kolb and D\"{u}nweg in section
\ref{sec:simulations}. For this purpose we present simulation results
of both a one-dimensional toy model that can be solved analytically,
and a three-dimensional Lennard-Jones system. We conclude the paper in
section \ref{sec:summary}.

\section{Isothermal-Isobaric Langevin Dynamics} 
\label{sec:method}

\subsection{Statistical mechanical considerations}

In his seminal paper on the extended Lagrangian formalism, Andersen
studied the statistical mechanics of $N$ particles within a box with a
fluctuating volume $V$ subject to a constant external pressure $P$
\cite{andersen}. He associated the volume fluctuations with the motion
of a ``piston'', and considered an extended phase space of $2(dN+1)$
degrees of freedom, including (i) the $Nd$ coordinates of the
particles, $\bar{r}_i$, and their $Nd$ conjugate momenta $\bar{p}_i$,
and (ii) the volume $V$ representing the coordinate of a ``piston''
along with its conjugate momentum. The derivation of the extended
Lagrangian formalism was done in the rather uncommon
isoenthalpic-isobaric ensemble $(N,P,H)$, where the enthalpy is
$H=E+PV$. This is the equivalent of the microcanonical ensemble
$(N,V,E)$ for fixed pressure.

The degrees of freedom $\bar{r}_i$ and $V$ in the extended system are
{\it not} independent of each other because the particle coordinates
are adjusted to volume fluctuations via simple scaling during MD
simulations. In order to have independent statistical variables, one
needs to define the scaled coordinates $\bar{s}_i$
\begin{eqnarray}
s_{i,\mu} & = & r_{i,\mu}/L_\mu ,\label{eq:Eq_Andersen_coordinate}
\end{eqnarray}
where $\mu=x,y,z$, $\bar{s}_i=(s_{i,x},s_{i,y},s_{i,z})^T$, and
$L_\mu$ is the linear size of the simulations box along the
$\mu$-axis. For simplicity, we here assume that the simulation box is
orthorhombic with fixed aspect ratios, such that $\prod_\mu L_\mu=V$,
and that all the particles have identical mass $m$, except for the
piston, which is considered a coordinate with inertial constant
$Q$. The ``force" acting on the piston is derived from the extended
Hamiltonian, which is obtained from the extended Lagrangian via
Legendre transformation. It is given by (see Eq.~(3.14C) in
Ref.~\cite{andersen})
\begin{equation}
f_P= \frac{1}{Vd}\sum_{i=1}^{N}
\left(\bar{f}_i\cdot\bar{r}_i+\frac{\bar{p}_i^2}{m}\right)-P.
\label{eq:presandersen}
\end{equation}

The transition from the isoenthalpic-isobaric into the
isothermal-isobaric ensemble requires the introduction of a
``thermostat'', and as noted in section \ref{sec:intro}, the
thermostat can be either ``deterministic'' (Nos\'{e}-Hoover) or
``stochastic'' (Langevin). In terms of the coordinates $\bar{s}_i$ and $V$,
the isothermal-isobaric partition function reads:
\begin{eqnarray}
&&Z=\int_0^{\infty}\!\!\!dV\,V^N \!\!\int_0^1
\mathop{\prod_{i=1}^{N}}_{\mu=x,y,z} ds_{i,\mu}
\,e^{-\left[U\left(\left\{L_\mu s_{i,\mu}\right\}\right)+PV\right]/k_BT}
\label{eq:npt-partition}
\\ &&=\int_0^{\infty}\!\!\!dV\!\!\int_0^1
\mathop{\prod_{i=1}^{N}}_{\mu =x,y,z}ds_{i,\mu}
\,e^{-\left[U\left(\left\{L_\mu s_{i,\mu}\right\}\right)+PV-Nk_BT\ln{V}\right]/k_BT}.
\nonumber
\end{eqnarray}
This partition function can be interpreted as if governing the
canonical ensemble of a system consisting of $N$ particles confined to
a three-dimensional unit cube ($0\leq s_{i,\mu}\leq 1$), and a piston
moving along an infinite line ($0<V<\infty$), with the potential
energy given by
\begin{equation}
U_{\rm eff}(\{s_{i,\mu}\},\{L_\mu\})=U(\{L_\mu s_{i,\mu}\})+PV-Nk_BT\ln{V}.
 \label{eq:ueff}
\end{equation} 
Notice that the partition function defined by
Eq.~(\ref{eq:npt-partition}) includes summation only over the spatial
degrees of freedom (of the particles and piston), but not over their
conjugate momenta. This deviation from Andersen's extended Lagrangian
formalism, where both the coordinates and momenta were included in the
partition sum, deserves an explanation. Andersen's method describes
Newtonian dynamics within a microcanonical ensemble. In this ensemble,
the kinetic and potential energies are coupled by energy
conservation. In contrast, Langevin dynamics occurs within an open
system in contact with a heat bath. In this canonical ensemble, the
kinetic and potential energies are decoupled, and the degrees of
freedom of the coordinates can be integrated separately from their
associated momenta. The momenta degrees of freedom follow a
Maxwell-Boltzmann Gaussian distribution, while the coordinates ($\bar{s}_i$
and $V$) are governed by the Boltzmann distribution corresponding to
$U_{\rm eff}$ (\ref{eq:ueff}). The separation of the ensemble into two
sub-spaces, corresponding to the coordinates and their associated
momenta, is important because {\em the goal of constant temperature
and pressure simulations is to sample the phase space of the
coordinates correctly.}\/ The momenta, i.e., the velocities, are only
used in these simulations as a mean to assess the simulated kinetic
temperature. The average kinetic energy is a reasonable measure of the
temperature, but not a good one in discrete-time because of the second
order (in $dt$) deviation between the measured velocity relative to
the trajectory of the corresponding coordinate (see Appendix
\ref{appendix_b}). Thus, numerical measures involving velocity are not
reliable for non-vanishing time steps. In constant volume simulations,
this problem is avoided if the Langevin dynamics is computed using the
accurate G-JF integrator, which exhibit no changes in the
configurational sampling statistics in response to variations in
$dt$. Moreover, the aforementioned closely-related problem of constant
pressure simulations resulting from the dependence of the ``piston
force'' on the velocities [see Eq.~(\ref{eq:presandersen})] is
eliminated as well, because, in the configuration phase space of
interest (which does not include momenta degrees of freedom), the
piston ``force" (pressure) is derived from $U_{\rm eff}$ (\ref{eq:ueff})
\begin{eqnarray}
f_P & = & -\frac{\partial U_{\rm eff}}{\partial V} \; = \;
\frac{1}{Vd}\sum_{i=1}^{N} \bar{f}_i\cdot\bar{r}_i+\frac{Nk_BT}{V}-P \nonumber \\ & =
& {\cal P} - P,
\label{eq:presgjf}
\end{eqnarray}
where we have defined the internal pressure ${\cal P}$.
This last important point was neither included by Andersen in his original
paper, nor in other later contributions on Langevin dynamics at constant
pressure.

\subsection{Derivation of the method}
\subsubsection{Dynamics of the volume}
Following Andersen's idea, we introduce the inertial coefficient $Q$
for a piston with a coordinate that coincides with the volume $V$ of
the system.  The ``regular'' force (pressure), $f_P$, acting on this
particle, is given by Eq.~(\ref{eq:presgjf}). The piston coordinate
moves with velocity ${\cal V}=\dot{V}$ in a medium with friction coefficient
$\tilde\alpha$ at constant temperature $T$. The Langevin dynamics of
this ``particle'' is
\begin{eqnarray}
Q\dot{{\cal V}}+\tilde{\alpha}{\cal V} & = & f_P+\tilde{\beta}(t).
\end{eqnarray}
This equation will be integrated using the G-JF thermostat, which
(in the velocity-Verlet form) is expressed by the following equations
to calculate the coordinate (i.e., volume) $V^{n+1}$ and velocity
${\cal V}^{n+1}$ at time $t_{n+1}=t_n+dt$ (See Eqs.~(4)-(8) in
Ref.~\cite{gjf2}):
\begin{eqnarray} 
\!\!\!\!\!V^{n+1}&=&V^n+\tilde{b}dt{\cal
V}^n+\frac{\tilde{b}dt^2}{2Q}f_P^n+\frac{\tilde{b}dt}{2Q}\tilde{\beta}^{n+1}
\label{eq:Eq_GJF_vol_q}\\ {\cal V}^{n+1}&=&\tilde{a}{\cal
V}^n+\frac{dt}{2Q}\left(\tilde{a}f_P^n+f_P^{n+1}\right)+\frac{\tilde{b}}{Q}\tilde{\beta}^{n+1}
\label{eq:Eq_GJF_vol_p},
\end{eqnarray}
where
\begin{eqnarray}
\tilde{a}=\frac{1-\frac{\tilde{a}dt}{2Q}}{1+\frac{\tilde{a}dt}{2Q}}\\
\tilde{b}=\frac{1}{1+\frac{\tilde{a}dt}{2Q}},
\end{eqnarray}
and $\tilde{\beta}^{n}$ is a normally distributed random number with
zero mean, 
and autocorrelation $\langle\tilde{\beta}^{n}\tilde{\beta}^m\rangle=2\tilde{\alpha}k_BTdt\delta_{m,n}$.

\subsubsection{Dynamics of the particles}
The variation of the volume causes complications for the dynamics of
the particles, which reside within the confines of the defined, yet
variable, volume. These complications are particularly apparent in
systems with periodic boundary conditions since the simulated volume
is associated with a lattice constant of a simulation box and not with
a physical location of an actual piston. Thus, in order to preserve
the translational invariance of the equations of motion in a bulk
system with periodic boundary conditions, it is necessary to globally
couple the dynamics of the volume to all the particles, regardless of
particle location in the simulation cell \cite{andersen}, such that
relative distances in the system are preserved. This is accomplished
through the scaled (normalized) coordinate
$s_{i,\mu}=r_{i,\mu}/L_\mu$, which is understood to be constant for a
simple expansion or contraction of $L_\mu$. However, the physical
velocity and acceleration of the coordinate $r_{i,\mu}$ can then not
be translational invariant without modifications.
Andersen's solution to the problem is to investigate the derivative
\begin{equation}
\dot{r}_{i,\mu}=\dot{s}_{i,\mu}L_\mu+s_{i,\mu}\dot{L}_\mu.
\label{eq:fullvelocity}
\end{equation}
In Eq.~(\ref{eq:fullvelocity}) we can see the separation of two
identifiable components to the motion: (i) the dynamics of the
particle relative to the simulation cell (first term on the r.h.s.),
and (ii) the dynamics due to the motion of the simulation cell (second
term on the r.h.s.). Thus, {\it defining}
\begin{eqnarray}
v_{i,\mu} & = & \dot{s}_{i,\mu}L_\mu 
\label{eq:Eq_vel_cont}
\end{eqnarray}
as the relevant {\em physical}\/ velocity of the coordinate
$r_{i,\mu}=s_{i,\mu}L_\mu$ makes the particle dynamics invariant to
the origin of the coordinate system -- an essential necessity for
meaningful dynamics. While this elegant observation has led to the
advanced formulations of both deterministic
\cite{nose,hoover,hoover2,klein,martyna} and stochastic
\cite{feller,kolb} methods for NPT simulations, the inherent problem
of time discretization persists.

We now arrive at the core of the derivation of the new method. We
simplify the notation for brevity in the rest of this subsection, such
that, e.g., the coordinate $r$ refers to $r_{i,\mu}$, $L$ to $L_\mu$,
etc, unless specifically indicated otherwise.

We reevaluate the particle equations of motion in discrete-time,
starting with the definition of the scaled coordinate
Eq.~(\ref{eq:Eq_Andersen_coordinate}). The total particle displacement
$\Delta r^{n+1}=r^{n+1}-r^n$ in one time step is given by
\begin{eqnarray}
&&\Delta{r}^{n+1}  =  \int_{t_n}^{t_{n+1}}\frac{d}{dt^\prime}(sL)\,dt^\prime \; = \; {s^{n+1}L^{n+1}-s^nL^n} \nonumber \\
\label{eq:Eq_r_integration}\\
&  & ={(s^{n+1}-s^n)\frac{L^{n+1}+L^n}{2}}+{\frac{s^{n+1}+s^n}{2}(L^{n+1}-L^n)} .\nonumber \\
\label{eq:Eq_r_partitioning}
\end{eqnarray}
We use the analogy between Eqs.~(\ref{eq:fullvelocity}) and
(\ref{eq:Eq_r_partitioning}) to {\it define} the relevant physical,
spatially invariant, discrete-time particle displacement $\Delta
q^{n+1}$ from the first term in Eq.~(\ref{eq:Eq_r_partitioning})
\begin{eqnarray}
\Delta q^{n+1} & = & (s^{n+1}-s^n)\frac{L^{n+1}+L^n}{2} .
\label{eq:Eq_def_del_q}
\end{eqnarray}
Notice that $\Delta q\rightarrow\dot{s}Ldt$ for $dt\rightarrow0$,
consistent with the usual continuous-time definition of the relevant
velocity mentioned above $v=\dot{s}L$ (\ref{eq:Eq_vel_cont}). Thus, we
conclude that the discrete-time particle dynamics must involve
the physical coordinate $q$ and an associated velocity $v$, which must
relate to the discrete-time displacement through
\begin{eqnarray}
\Delta q^{n+1}=\int_{t_n}^{t_{n+1}}v\, dt^\prime.
\label{eq:Eq_phys_q_v}
\end{eqnarray}
The corresponding discrete-time velocity change $\Delta
v^{n+1}=v^{n+1}-v^n$ is obtained through the $dt$-integrated Langevin
equation
\begin{eqnarray}
\int_{t_n}^{t_{n+1}}\left[m\dot{v}+\alpha v\right]\, dt^\prime
=\int_{t_n}^{t_{n+1}}\left[f(r,t^\prime)+\beta(t^\prime)\right]\, dt^\prime,
\end{eqnarray}
which, using (\ref{eq:Eq_phys_q_v}) and with no approximation, can be written 
\begin{eqnarray}
m\Delta{v}^{n+1}+\alpha\Delta{q}^{n+1} & = & \int_{t_n}^{t_{n+1}}f\,
dt^\prime + \beta^{n+1} \; , \label{eq:Eq_GJF_pos_0}
\end{eqnarray}
where we have defined the Wiener process
\begin{eqnarray}
\beta^{n+1} & = & \int_{t_n}^{t_{n+1}}\beta(t^\prime) \, dt^\prime
\end{eqnarray}
such that
\begin{eqnarray}
\langle\beta^n\rangle & = & 0 \; \; , \; \; \; \langle\beta^n\beta^m\rangle \; = \; 2\alpha k_BTdt \delta_{n,m}\; .
\end{eqnarray}
(The noise autocorrelation reads with full notation:
$\langle\beta_{i,\mu}^n\beta_{j,\nu}^m\rangle \; = \; 2\alpha k_BTdt
\delta_{n,m}\delta_{i,j}\delta_{\mu,\nu}$.)  Notice that the
introduction of the discrete-time Langevin equation in
Eq.~(\ref{eq:Eq_GJF_pos_0}), for linking the coordinate displacement
$\Delta q$ with its velocity change $\Delta v$, ensures physically
meaningful discrete-time evolution.

Starting from Eq.~(\ref{eq:Eq_phys_q_v}), and following our previous
work \cite{gjf}, we now choose the time-reversible relationship
between the relative displacement and change in the associated
velocity:
\begin{eqnarray}
\Delta{q}^{n+1} & = & \int_{t_n}^{t_{n+1}}v\, dt^\prime \; \approx \; \frac{dt}{2}(v^{n+1}+v^n) \nonumber \\
& = & \frac{dt}{2}\Delta{v}^{n+1}+dtv^n . \label{eq:Eq_GJF_pos_1}
\end{eqnarray}
Inserting (\ref{eq:Eq_GJF_pos_0}) into (\ref{eq:Eq_GJF_pos_1}) yields
\begin{eqnarray}
\Delta{q}^{n+1} & = & b\,dt\,v^n+\frac{b\,dt}{2m}\int_{t_n}^{t_{n+1}}f\, dt^\prime + \frac{b\,dt}{2m}\beta^{n+1} \label{eq:Eq_GJF_q}
\end{eqnarray}
where
\begin{eqnarray}
b & = & \frac{1}{1+\frac{\alpha dt}{2m}} \; .
\end{eqnarray}
Equations (\ref{eq:Eq_GJF_q}) and (\ref{eq:Eq_GJF_pos_0}) constitute a
set of equations for determining $\Delta q^{n+1}$ and $\Delta
v^{n+1}$.  We then approximate the $dt$-integrals over the
deterministic force $f$ such that all terms in the equations become at
least second order correct in $dt$ (i.e., consistent with the
traditional Verlet methods), which yields:
\begin{eqnarray}
\Delta{q}^{n+1} & = & b\,dt\,v^n+\frac{b\,dt^2}{2m}f^n + \frac{b\,dt}{2m}\beta^{n+1} \label{eq:Eq_gjf_q}\\
\Delta{v}^{n+1} & = & -\frac{\alpha}{m}\Delta{q}^{n+1} +  \frac{dt}{2m}(f^n+f^{n+1})+ \frac{1}{m}\beta^{n+1} \nonumber \\ \label{eq:Eq_gjf_vq}
\end{eqnarray}
These are explicit discrete-time equations for evaluating the
evolution of the coordinates $q^n$ and $v^n$.

In order to express the equations in the most useful form for
molecular simulations, we use the relationship $r^n=s^nL^n$
(Eq.~(\ref{eq:Eq_Andersen_coordinate})) to combine
Eqs.~(\ref{eq:Eq_def_del_q}) and (\ref{eq:Eq_gjf_q}) for a direct
expression of the dynamics of the physical coordinate $r^n$:
\begin{eqnarray}
r^{n+1} & = & \frac{L^{n+1}}{L^n}r^n\nonumber \\
&+&\frac{2L^{n+1}}{L^{n+1}+L^n}\,b\,dt\,[v^n+\frac{dt}{2m} f^n + \frac{1}{2m}\beta^{n+1}] \nonumber \\
\label{eq:Eq_gjf_bar_r}
\end{eqnarray}
We also insert Eq.~(\ref{eq:Eq_gjf_q}) into Eq.~(\ref{eq:Eq_gjf_vq})
in order to obtain an explicit equation for the dynamics of the
velocity $v^n$:
\begin{eqnarray}
v^{n+1} & = & av^n+\frac{dt}{2m}(af^n+f^{n+1})+\frac{b}{m}\beta^{n+1} \label{eq:Eq_gjf_bar_v}\; ,
\end{eqnarray}
where
\begin{eqnarray}
a & = & \frac{1-\frac{\alpha dt}{2m}}{1+\frac{\alpha dt}{2m}}
\end{eqnarray}
Equations (\ref{eq:Eq_gjf_bar_r}) and (\ref{eq:Eq_gjf_bar_v}) are the
Verlet-type equations for particle updates in the stochastic G-JF
thermostat/barostat, given a change $\Delta L^{n+1}=L^{n+1}-L^n$ in
simulation box dimension during the time step $dt=t_{n+1}-t_n$. Notice
that the velocity equation depends only indirectly on the change in
the simulation dimension $L$ through the force
$f^n=f(r^n,t_n,L^n)$. 

To summarize, given $r^n$, $v^n$, $f^n$, $f_P^n$, $V^n$, and ${\cal
V}^n$, the discrete-time dynamics evolves according to the following
protocol:

\noindent 1. Compute $V^{n+1}$ (and $L_\mu^{n+1}$) using Eq.~(\ref{eq:Eq_GJF_vol_q}).

\noindent 2. Compute $r^{n+1}$ (all $\bar{r}_i^{n+1}$) using
Eq.~(\ref{eq:Eq_gjf_bar_r}).

\noindent 3. Evaluate the new forces
$f^{n+1}=f(r^{n+1},t_{n+1},L^{n+1})$ (all $\bar{f}_i^{n+1}$), and
$f_P(\{\bar{r}_i^{n+1}\},t_{n+1},\{L_\mu^{n+1}\})$.
 
\noindent 4. Compute ${\cal V}^{n+1}$ using
Eq.~(\ref{eq:Eq_GJF_vol_p}) and $v^{n+1}$ (all $\bar{v}_i^{n+1}$)
using Eq.~(\ref{eq:Eq_gjf_bar_v}).\ \ \ \ \

\noindent We reemphasize that the coordinates $r^n$, $v^n$, and $f^n$
here refer to each Cartesian coordinate of each particle, and that $L$
refers to $L_\mu$, such that $V=\prod_\mu L_\mu$ is the volume of an
isotropically varying, orthorhombic simulation box.

\section{Testing the Algorithm}
\label{sec:simulations}

In order to test the method we have applied it to two characteristic
systems with a specific eye on the robustness against time-step
variations. We compare our results with those generated by the KD
method \cite{kolb}, which represents state-of-the-art of a sound
approach to Langevin dynamics NPT simulations in atomic and molecular
ensembles. The first system is a particular non-trivial,
one-dimensional particle model for which we can analytically derive
measurable thermodynamic quantities. This model therefore serves as a
strict benchmark for the statistical accuracy of a numerical test
simulation. The second system is the foundational model system in
computational statistical mechanics, namely the three-dimensional
ensemble of particles interacting with a Lennard-Jones force field. In
this latter case, we do not have analytical expressions for the
statistical measures, but we investigate the measures for different
values of the discrete time step, and from that infer the quality of
the applied numerical methods.

\subsection{One-dimensional model system}
\label{sec:model_1D}
We consider a one dimensional system of normalized length $L$
(characteristic length $r_0$) with periodic boundary conditions. $N$
identical particles are located in order at $\{x_1,x_2,\ldots,x_N\ ;\
x_i<x_{i+1}\}$ such that the periodic boundary conditions ensure two
neighbors for each particle (i.e., $x_{i\pm N}=x_i\pm L$). Each
particle interacts with its two neighbors via a pair-potential that
depends on the normalized pair distance $r$. Expressing the energy in
units of the thermal energy ($E_0=k_BT$), the normalized pair-potential
$u(r)$ [related to the physical potential via $U(r_0r)=E_0u(r)$] reads
\begin{eqnarray}
u(r)=\frac{\epsilon}{r}+\frac{1}{2}\ln{r}.
\end{eqnarray}

The pair potential $u(r)$ consists of two contributions: a repulsive
part ($\epsilon>0$), inversely proportional to $r$, and an attractive
logarithmic part. The latter may represent an entropic potential of
mean force resulting from implicit degrees of freedom. Considering the
isobaric-isothermal ensemble $(N,P,T)$ (where $P$ denotes the
one-dimensional pressure, i.e., the force, and is expressed in units
of $E_0/r_0$), the partition function of the system is given by
\begin{eqnarray}
Z=\int dL\prod_{i=1}^N\int dx_i 
\exp\left[-\sum_{i=1}^Nu\left(x_{i+1}-x_i\right)-PL\right].
\end{eqnarray}
Switching to the set of variables $r_i=x_{i+1}-x_i$, the partition
function reads 
\begin{eqnarray} 
Z&=&\prod_{i=1}^N \int_0^{\infty}dr_i
\exp\left[-\sum_{i=1}^Nu\left(r_i\right)-P\sum_{i=1}^N r_i\right]
\nonumber\\
&=& \left[\int_0^{\infty}\frac{dr}{\sqrt{r}}\,e^
{-\left(\epsilon/r+Pr\right)}\right]^N
\nonumber \\
&=& \left[2\int_0^{\infty}dy\,e^
{-\left(\epsilon/y^2+Py^2\right)}\right]^N,
\end{eqnarray}
where the last equality has been obtained by setting $y=\sqrt{r}$. The
value of the last integral is known, giving
\begin{eqnarray}
Z=\left[\sqrt{\frac{\pi}{P}}\,e^{-2\sqrt{P\epsilon}}\right]^N.
\end{eqnarray}
The normalized Gibbs free energy is given by $G=-\ln Z$, and the mean
nearest neighbor particle normalized distance, $\langle l\rangle $, is
then derived by
\begin{eqnarray}
\langle l \rangle \equiv\frac{\langle L\rangle}{N}=-\frac{1}{N}\frac{\partial G}{\partial P}
=\frac{1}{2P}
+\sqrt{\frac{\epsilon }{P}}.
\label{eq:first}
\end{eqnarray}
The variance of the normalized length distribution is given by
\begin{eqnarray}
\sigma_l^2 & \equiv & \frac{\left\langle(L-\langle L\rangle)^2\right\rangle}{N}\nonumber \\
&=&
-\frac{\partial\langle l\rangle}{\partial P}=\frac{1}{2P^2}+
\frac{\sqrt{\epsilon }}{2}P^{-\frac{3}{2}}.
\label{eq:second}
\end{eqnarray}

\begin{figure}[t]
\begin{flushleft}
\scalebox{0.55}{\centering \includegraphics{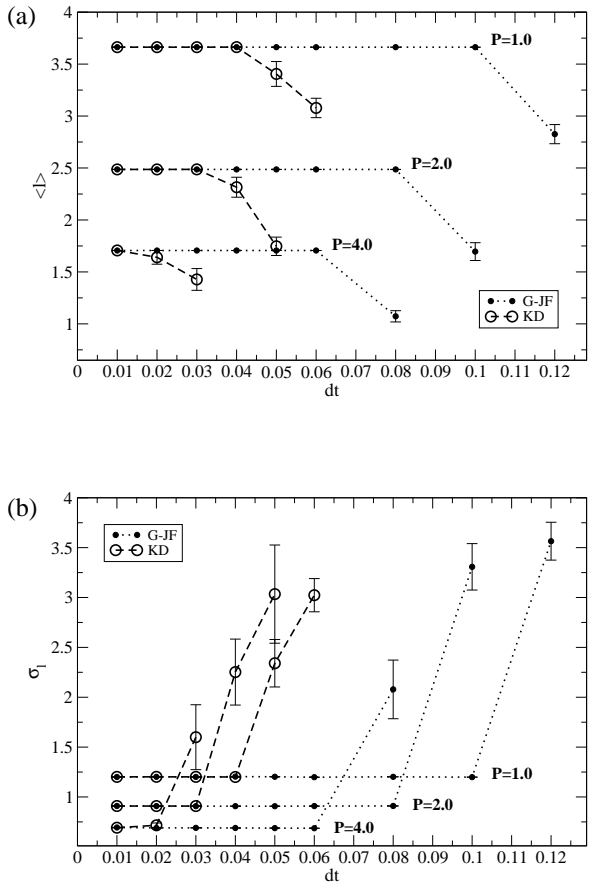}}
\end{flushleft}
\vspace{-0.75cm}
\caption{Results for $\epsilon=10$. Simulated average length (a) and
standard deviation $\sigma_l$ (b) for several values of applied 1D
pressure (force) $P$. Markers represent the G-JF method of this paper
(solid marker $\bullet$) and the KD method (open marker $\circ$). At
small time steps both methods produce the correct analytical values
given by Eqs.~(\ref{eq:first}) and (\ref{eq:second}). All simulations
were done with $Q=10^{-2}$ and $\tilde{\alpha}=10^{-2}$. Lines serve
as guides to the eye.}
\label{fig:fig_OA}
\end{figure}

\begin{figure}[h!]
\begin{flushleft}
\scalebox{0.3}{\centering \includegraphics{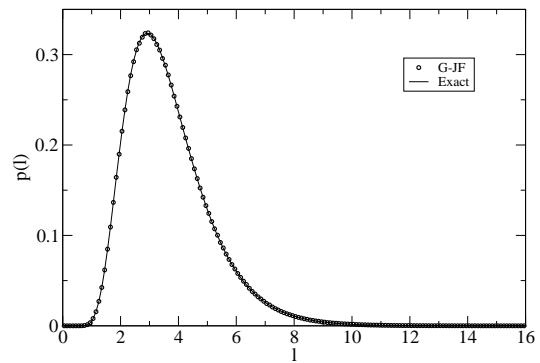}}
\end{flushleft}
\vspace{-0.75cm}
\caption{Results for $\epsilon=10$, $P=1$ and $dt=0.06$. The computed
length distribution (open circles) compared with the analytically
calculated exact distribution (solid curve).}
\label{fig:hist}
\end{figure}

For a system of $N=1000$ particles, we simulate the evolution for a
normalized transient time of $4.8\times 10^4$ units before producing
statistical averages of $\langle l \rangle$ and $\sigma_l$ over the
next $4.8\times 10^6$ normalized time units. Figures
\ref{fig:fig_OA}(a) and (b) show the resulting data for both the G-JF
method of this paper (solid markers, dotted line) and the KD method
(open markers, dashed line) for three different values of the
external, one-dimensional pressure $P$, with particle mass and
dissipation normalized parameters $m=1$ and $\alpha=1$,
respectively. By inspecting the convergence of $l$ to its equilibrium
value, we can find values for the normalized piston parameters that
provide efficient relaxation. For the model system discussed herein,
we choose $Q=\tilde{\alpha}=10^{-2}$. The acquired data clearly show
that the G-JF method is extremely accurate. The computed values of
both the average and fluctuations of the length agree with the
predictions of Eqs.~(\ref{eq:first}) and (\ref{eq:second}). The
accuracy of the method is also demonstrated in Fig.~\ref{fig:hist},
where the full length distribution $p(l)$ is plotted for $P=1$ and
$dt=0.06$. The agreement with the analytically calculated exact
distribution is perfect. Another important feature of the method,
demonstrated in Figure \ref{fig:fig_OA}, is its robustness against
time step variations. In comparison, we observes in
Fig.~\ref{fig:fig_OA} that the KD method yields the correct result for
small $dt$, but that the stability range is generally considerably
smaller than for the G-JF procedure.  In a different set of
simulations (data not shown), we used $Q=10^{-4}$ and
$\tilde{\alpha}=0$. This choice of parameters made the KD barostat
unstable for all the simulated values of $dt$ ($dt\geq 0.01$), while
keeping almost unchanged the stability range of the G-JF method. The
relative robustness of the latter against variations in the piston
parameters is yet another merit of this method.

\subsection{Three-dimensional Lennard-Jones model system}
\begin{figure}[t]
\begin{center}
\scalebox{0.5}{\centering
\includegraphics{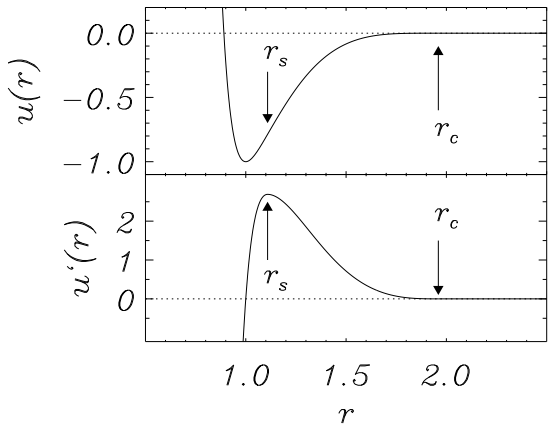}}
\end{center}
\vspace{-0.5cm}
\caption{Particle interaction as given in
Eq.~(\ref{eq:Eq_LJ_spline}). Upper plot shows $u(r)$, lower plot shows
$u^\prime(r)$. Spline point and cut-off distance are indicated by
arrows.}
\label{fig:fig_A}
\end{figure}

We now consider the simplest possible well-known system in the
modeling of materials and liquids, namely a three-dimensional ensemble
of identical spherical particles. Each particle has a normalized mass
$m=1$ (in units of $m_0$) and normalized friction coefficient
$\alpha=1$, and they all interact through the normalized potential
$u(r)$ given by the physical pair-potential $U(\{r_0r\})=E_0 u(r)$,
where $r=|\bar{r}|$ is the normalized pair-distance (in units if the
characteristic length $r_0$) and $E_0$ is the characteristic
energy. The normalized pair-potential reads
\begin{eqnarray}
u(r) & = & \left\{\begin{array}{lcl} r^{-12}-2r^{-6} & , & 0<r\le r_s
\\ a_4(r-r_c)^4+a_8(r-r_c)^8 & , & r_s< r<r_c \\ 0 & , & r_c \le r
\end{array}\right . \nonumber \\ \label{eq:Eq_LJ_spline}
\end{eqnarray}
where
\begin{eqnarray}
r_s & = & \left(\frac{13}{7}\right)^{1/6} \; \approx \; 1.108683 \\
r_c & = & r_s-\frac{32u(r_s)}{11u^\prime(r_s)} \; \approx \; 1.959794\\
a_4 & = & \frac{8u(r_s)+(r_c-r_s)u^\prime(r_s)}{4(r_c-r_s)^4} \\
a_8 & = & -\frac{4u(r_s)+(r_c-r_s)u^\prime(r_s)}{4(r_c-r_s)^8} .
\end{eqnarray}
This function (see Fig.~\ref{fig:fig_A}) is a short-range splined
Lennard-Jones potential with continuity through the second derivative
at $r=r_s$ and continuity through third derivative at $r=r_c$.

Conducting NPT simulations on a cubic system with $N=864$ particles,
we optimize the relaxation of the barostat degree of freedom $V$ by
choosing small values for the inertia $Q$. By inspection, we find that
values in the range $Q=10^{-4}$ and $Q=10^{-5}$ represent efficient
relaxation. We also, by inspection, conclude that a small friction
coefficient $\tilde{\alpha}=10^{-4}$ helps relax the system (although
this seems to be a weak effect) and, therefore, choose this value for
our simulations. We have further chosen two characteristic normalized
temperatures, for both solid ($k_BT/E_0=0.3$) and liquid
($k_BT/E_0=0.7$) phases. Finally, we have studied three different
applied pressures ($P=0.01, 0.1, 1.0$) (expressed in units of
$E_0/r_0^3$), and varied the discrete normalized time step $dt$ in the
entire range of stability to observe the behavior of the numerical
methods. We only show the $P=0.1$ data here since the results of all
three applied pressures exhibit the same characteristics.  All
statistical data are obtained by initiating the system in a
close-packed crystal near a zero-temperature ground state. We then
simulate at least 2$\times$10$^{5}$ normalized time units before
averages are acquired over the next 2$\times$10$^{5}$ units. The
normalizing time $\tau_0$ is given by $E_0\tau_0^2=m_0r_0^2$. All the
left axes of the figures display absolute results, while all the right
vertical axes display the percentage deviation from the
$dt\rightarrow0$ value of the quantity shown in the plot.

\begin{figure}[t]
\begin{flushleft}
\scalebox{0.37}{\centering \includegraphics{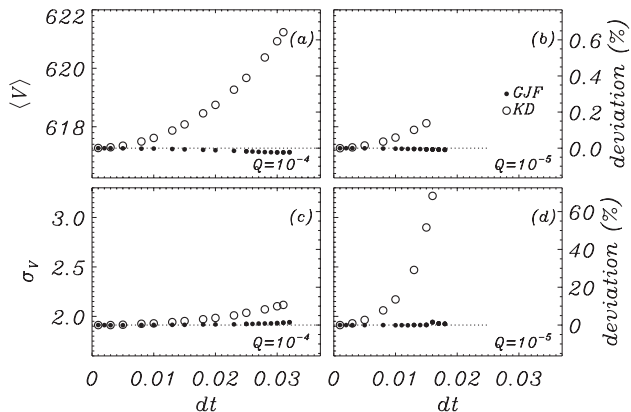}}
\end{flushleft}
\vspace{-0.5cm}
\caption{For $k_BT/E_0=0.3$ (solid phase): Simulated average volume
$\langle{V}\rangle$ [(a) and (b)] and standard deviation $\sigma_{V}$
[(c) and (d)] for $Q=10^{-4}$ [(a) and (c)] and $Q=10^{-5}$ [(b) and
(d)]. Markers represent the G-JF method of this paper (solid marker
$\bullet$) and the KD method (open marker $\circ$). Horizontal dotted
lines are leveled at $\langle{V}\rangle$ for $Q=10^{-4}$ and
$dt=0.001$ [(a) and (b)], and at $\sigma_{V}$ for $Q=10^{-4}$ and
$dt=0.001$ [(c) and (d)]. All figures show axes with absolute
quantities on the left and percentage deviation on the right axes.}
\label{fig:fig_2V}
\end{figure}

Figure \ref{fig:fig_2V} shows the data for the volume $V$ of the
simulation box ($Q=10^{-4}$: Fig.~\ref{fig:fig_2V}a; $Q=10^{-5}$:
Fig.~\ref{fig:fig_2V}b) and its fluctuations ($Q=10^{-4}$:
Fig.~\ref{fig:fig_2V}c; $Q=10^{-5}$: Fig.~\ref{fig:fig_2V}d) as a
function of the time step for a solid phase at $k_BT/E_0=0.3$ and
external pressure $P=0.1$. The new G-JF barostat results are displayed
as solid markers ($\bullet$), while the comparison KD method results
are shown with open markers ($\circ$).  The data clearly shows that
the G-JF results are nearly independent of the time step $dt$ for both
the average volume and the corresponding fluctuations.  In comparison,
the KD method exhibits a consistent, albeit weak, increase in average
volume.  More dramatic is the increasing deviation of the volume
fluctuations in the KD method. For $Q=10^{-4}$, this can be in excess
of 10\%, while we observe up to 70\% discrepancy for $Q=10^{-5}$. Such
discrepancies clearly change not only measured thermodynamic
properties such as the elastic bulk modulus and heat expansion
coefficient, but also the structure of the material under
investigation.  For example, close inspection (not shown) of the KD
simulation shows that the excessive volume fluctuations induce crystal
defects into the material for large $dt>0.016$, before the numerical
instability is found for $dt\approx0.019$. Notice that the results of
both methods converge to the same numbers for small $dt$ throughout
the simulation data, indicating that any deviation from small $dt$
constitutes a measure of the error induced exclusively by the discrete
time step.

\begin{figure}[t]
\begin{flushleft}
\scalebox{0.37}{\centering \includegraphics{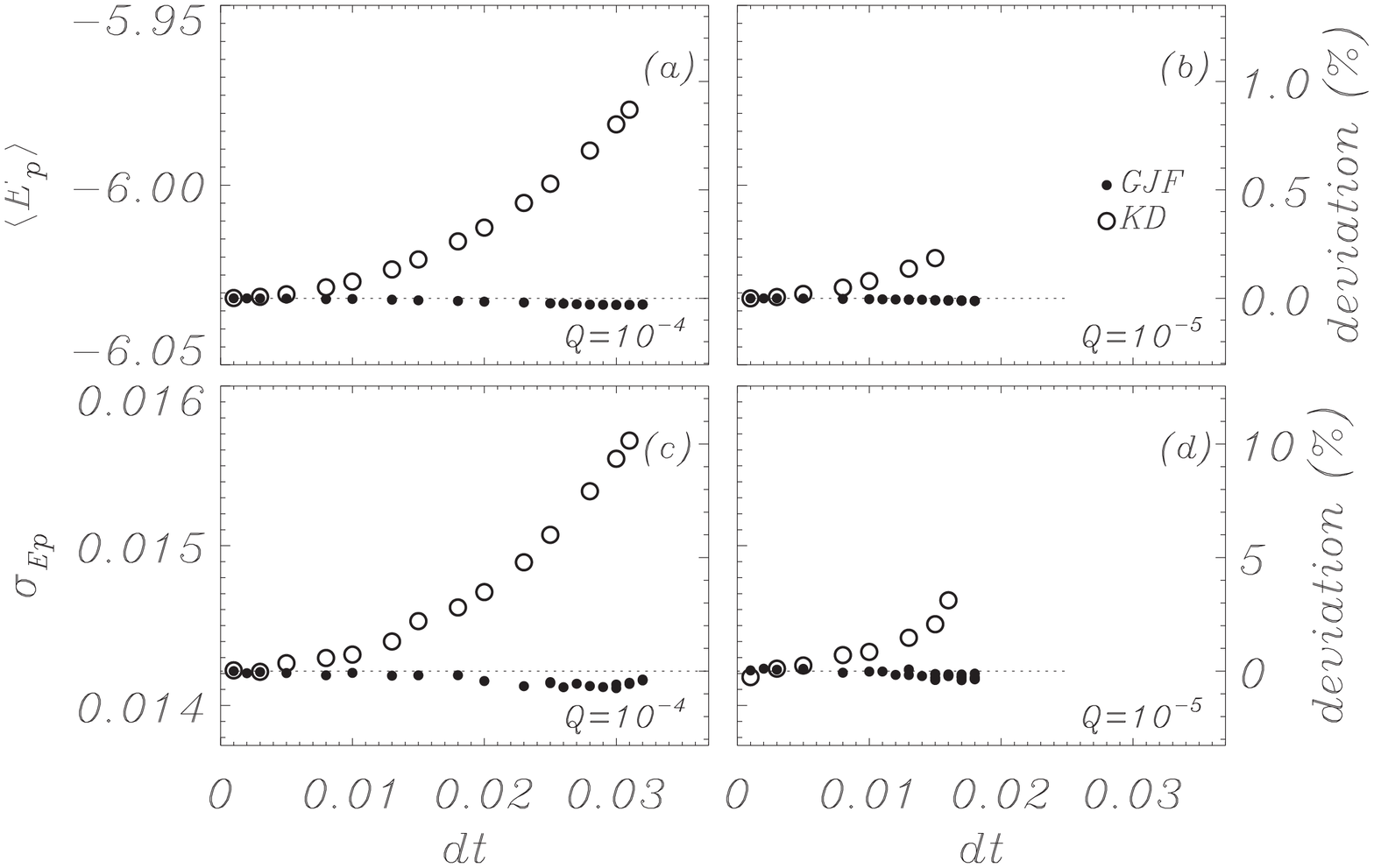}}
\end{flushleft}
\vspace{-0.5cm}
\caption{For $k_BT/E_0=0.3$ (solid phase): Simulated average potential
energy $\langle{E_p}\rangle$ [(a) and (b)] and standard deviation
$\sigma_{E_p}$ [(c) and (d)] for $Q=10^{-4}$ [(a) and (c)] and
$Q=10^{-5}$ [(b) and (d)]. Markers represent the G-JF method of this
paper (solid $\bullet$) and the KD method (open $\circ$). Horizontal
dotted lines are leveled at $\langle{E_p}\rangle$ for $Q=10^{-4}$ and
$dt=0.001$[(a) and (b)], and at $\sigma_{E_p}$ for $Q=10^{-4}$ and
$dt=0.001$ [(c) and (d)]. All figures show axes with absolute
quantities on the left and percentage deviation on the right axes.}
\label{fig:fig_2pot}
\end{figure}

\begin{figure}[t]
\begin{flushleft}
\scalebox{0.37}{\centering \includegraphics{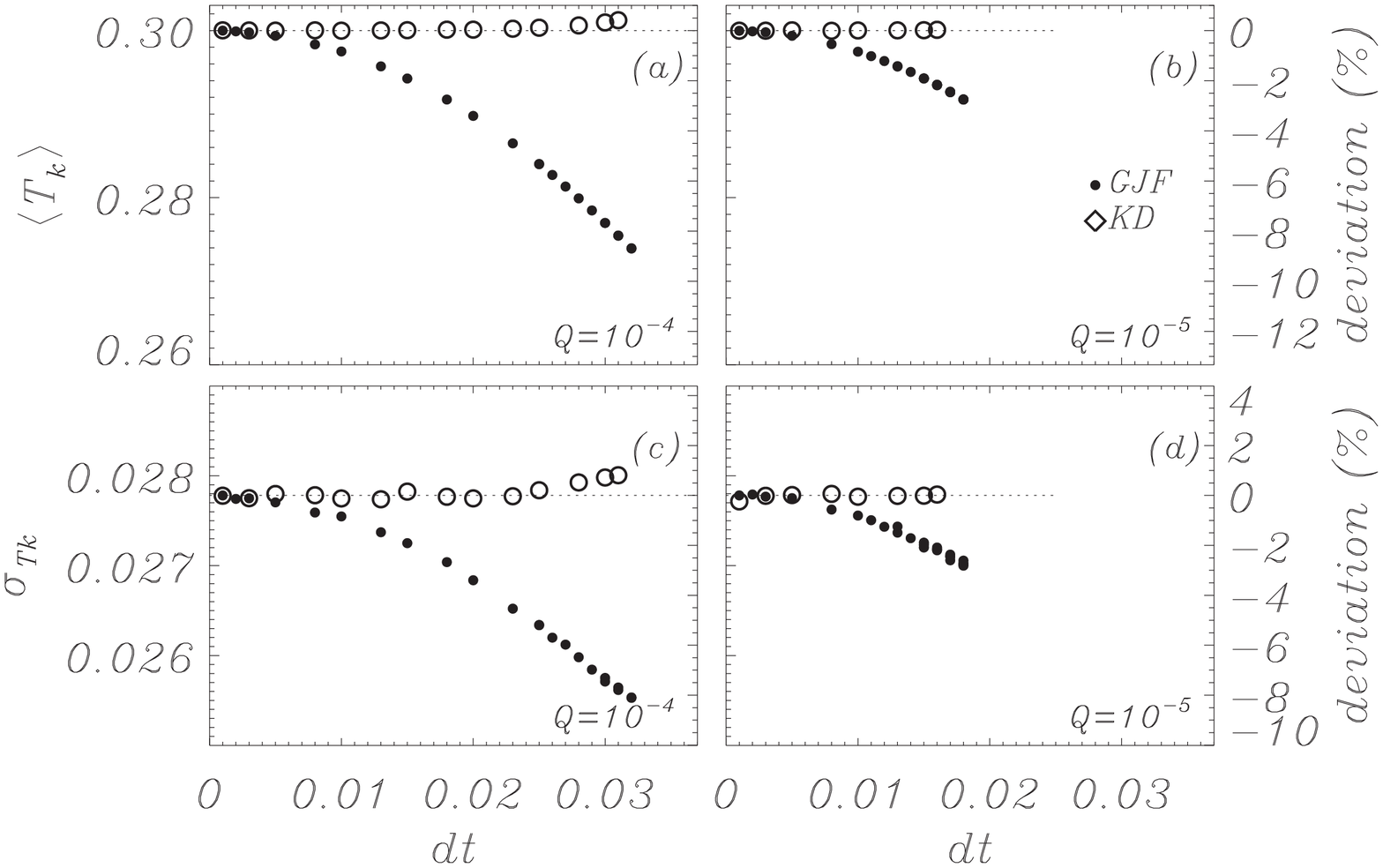}}
\end{flushleft}
\vspace{-0.5cm}
\caption{For $k_BT/E_0=0.3$ (solid phase): Simulated average kinetic
temperature $\langle{T_k}\rangle$ [(a) and (b)] (from
Eq.~(\ref{eq:Eq_Tk})) and standard deviation $\sigma_{T_k}$ [(c) and
(d)] for $Q=10^{-4}$ [(a) and (c)] and $Q=10^{-5}$ [(b) and
(d)]. Markers represent the G-JF method of this paper (solid
$\bullet$) and the KD method (open $\circ$). Horizontal dotted lines
are leveled at $\langle{T_k}\rangle=0.3$ for [(a) and (b)], and at
$\sigma_{T_k}$ for $Q=10^{-4}$ and $dt=0.001$ [(c) and (d)]. All
figures show axes with absolute quantities on the left and percentage
deviation on the right axes.}
\label{fig:fig_2kin}
\end{figure}

The data for the total potential energy 
\begin{eqnarray}
E_p=\sum_{i<j}u(r_{ij})+PV
\end{eqnarray}
are shown in Fig.~\ref{fig:fig_2pot}. The results of the G-JF method
are also here unimpressed with the simulated time step throughout the
stability ranges, while the KD method shows a characteristic positive
deviation.  Since the KD method coincides with the BBK thermostat
\cite{bbk} when the volume is constant, this result is entirely
expected in light of our previous work on the G-JF thermostat and
comparisons \cite{gjf2} to other thermostats, including BBK. These
results are also consistent with the kinetic temperature $T_k$
measurements shown in Figure \ref{fig:fig_2kin}, where $T_k$ is
defined as
\begin{eqnarray}
T_k & = & \frac{1}{3Nk_B}\sum_{i=1}^Nm\left(\bar{v}_i^n\right)^2
. \label{eq:Eq_Tk}
\end{eqnarray}
In the G-JF method, the kinetic temperature decreases with increasing
$dt$. This is anticipated since, as mentioned in the introduction, it
is known that the momentum $mv_i^n$ is {\it not} the conjugate
variable to $r_i^n$ for $dt>0$ (see, e.g.,
Refs.~\cite{eastwood,gjf,gjf2} as well as Appendix
\ref{appendix_b} below), and that the discrete-time second order
approximations ($v_i^n$ and ${\cal V}^n$) to the velocity variables
from the central difference approach in the Verlet formalism leaves
kinetic and configurational measures mutually inconsistent. Thus, the
very good configurational sampling properties of the G-JF method seen
from the measurements of, e.g., volume and enthalpy (along with their
fluctuations) inevitably mean that a measurement [such as $T_k$ in
Eq.~(\ref{eq:Eq_Tk})] derived from the (incorrect) velocities will be
incorrect to second order in $dt$. This is a simple consequence of
$v_i^n$ being an {\it approximation} to the true velocity of $r_i^n$,
which cannot be obtained. Figure \ref{fig:fig_2kin} further displays
the measured kinetic temperature $T_k$ for the KD method, and it is
apparent that this quantity appears to confirm the required
temperature, which is consistent with the {\em incorrect}\/
configurational properties of this method seen for volume and enthalpy
as $dt$ is increased. This artifact of discrete time emphasizes that
one should refrain from using kinetic measures as reliable quantities
in these types of simulations.

\begin{figure}[t]
\begin{flushleft}
\scalebox{0.37}{\centering \includegraphics{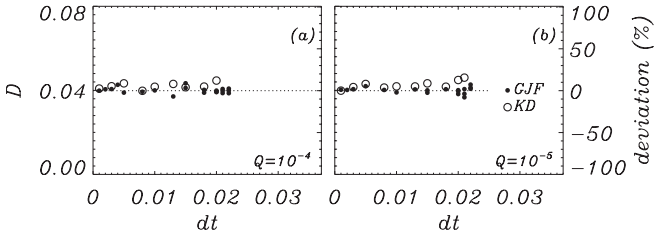}}
\end{flushleft}
\vspace{-3.0cm}
\caption{For $k_BT/E_0=0.7$ (liquid phase): Simulated diffusion
coefficient $D$ from Eq.~(\ref{eq:Eq_diffusion}) for $Q=10^{-4}$ (a)
and $Q=10^{-5}$ (b). Markers represent the G-JF method of this paper
(solid $\bullet$) and KD method (open $\circ$). Horizontal dotted
lines are leveled at $D$ for $Q=10^{-4}$ and $dt=0.001$. Both figures
show axes with absolute quantities on the left and percentage
deviation on the right axes.}
\label{fig:fig_3D}
\end{figure}

We now show results for a liquid phase at $k_BT/E_0=0.7$.  Otherwise,
all system and simulation parameters are exactly as for the
$k_BT/E_0=0.3$ results shown above. The liquid phase is validated by
structural analysis and through the measured diffusion constant, which
we derive from the Einstein definition
\begin{eqnarray}
D & = &
\frac{1}{N}\sum_{i=1}^N\lim_{ndt\rightarrow\infty}\left\langle{V}\right\rangle^{\frac{2}{3}}\frac{(\bar{s}_i^n-\bar{s}_i^0)^2}{6ndt}
\label{eq:Eq_diffusion}.
\end{eqnarray}
We use time averages over $ndt=200,000$ for all chosen values of $dt$,
and $\bar{s}_i^n$ is understood to extend beyond the interval
$0\le{s}<1$ in this expression \cite{velocity_diff}.  Figure
\ref{fig:fig_3D} displays the non-zero measured diffusion coefficient of the
liquid state as a function of the time step. It is clear that the
migration at this temperature and pressure is weak, and that the
diffusion measurement is noisy. Even so, the figure demonstrates that
both G-JF and KD methods exhibit diffusion coefficients reasonably
independent of the choice of the size of time step, although there may
be a hint of a slight increase for the KD method for increasing $dt$.

\begin{figure}[t]
\begin{flushleft}
\scalebox{0.37}{\centering \includegraphics{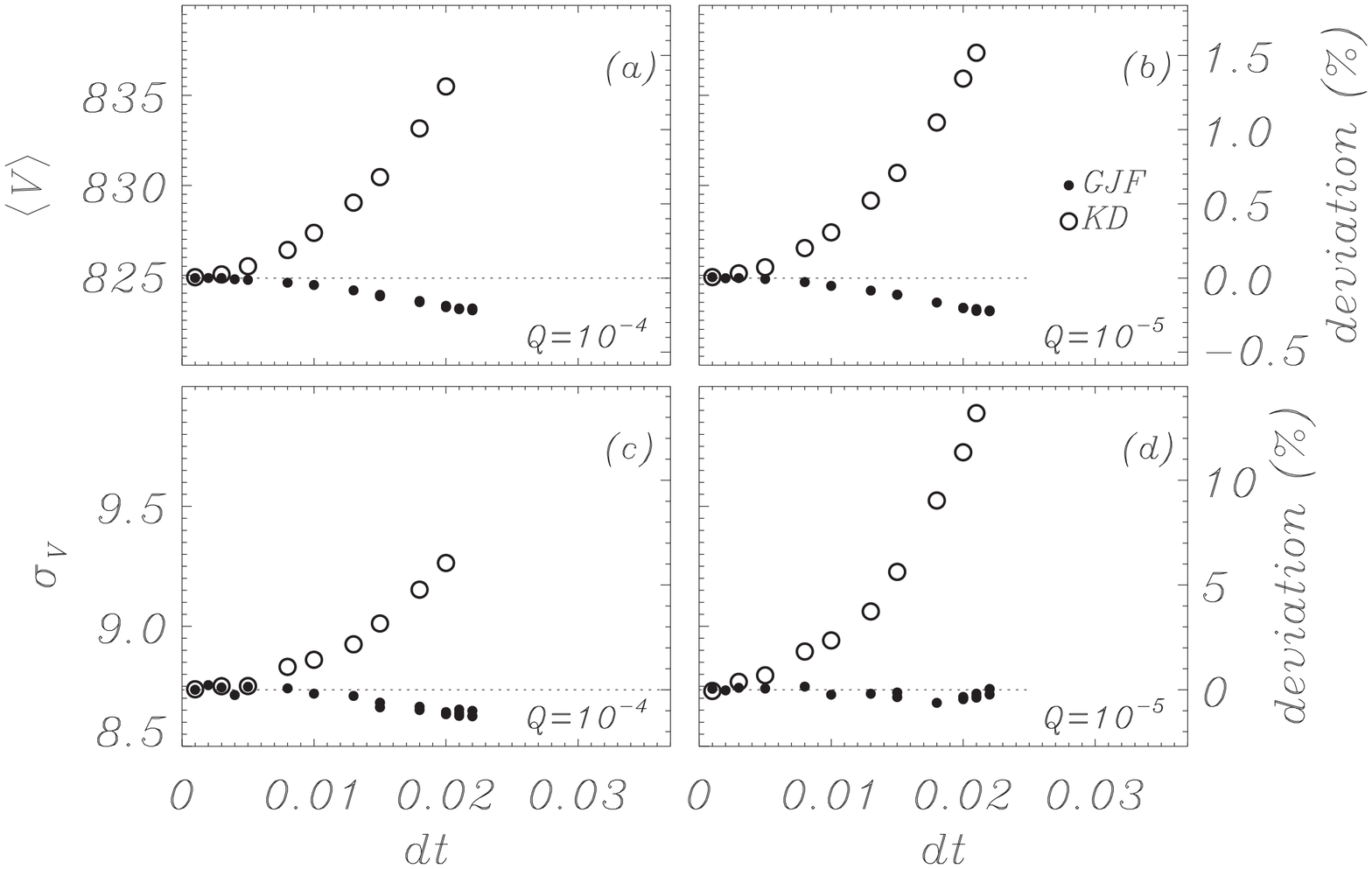}}
\end{flushleft}
\vspace{-0.5cm}
\caption{For $k_BT/E_0=0.7$ (liquid phase): Simulated average volume
$\langle{V}\rangle$ [(a) and (b)] and standard deviation $\sigma_{V}$
[(c) and (d)] for $Q=10^{-4}$ [(a) and (c)] and $Q=10^{-5}$ [(b) and
(d)]. Markers represent the G-JF method of this paper (solid
$\bullet$) and the KD method (open $\circ$). Horizontal dotted lines
are leveled at $\langle{V}\rangle$ for $Q=10^{-4}$ and $dt=0.001$ [(a)
and (b)], and at $\sigma_{V}$ for $Q=10^{-4}$ and $dt=0.001$ [(c) and
(d)]. All figures show axes with absolute quantities on the left and
percentage deviation on the right axes.}
\label{fig:fig_3V}
\end{figure}

Figure \ref{fig:fig_3V} shows the $k_BT/E_0=0.7$ data for the volume
$V$ of the simulation box.  The KD results for both average and
fluctuation of the volumes exhibit significant increases with $dt$,
consistent with the comparable $k_BT/E_0=0.3$ data. The G-JF results
are much less impressed by the time step $dt$, but there is a small
tendency for the volume and its fluctuations to decrease with
increasing $dt$. However, the overall impression is clearly that the
G-JF method is significantly less dependent on variations in $dt$ than
the KD method is.

\begin{figure}[t]
\begin{flushleft}
\scalebox{0.37}{\centering \includegraphics{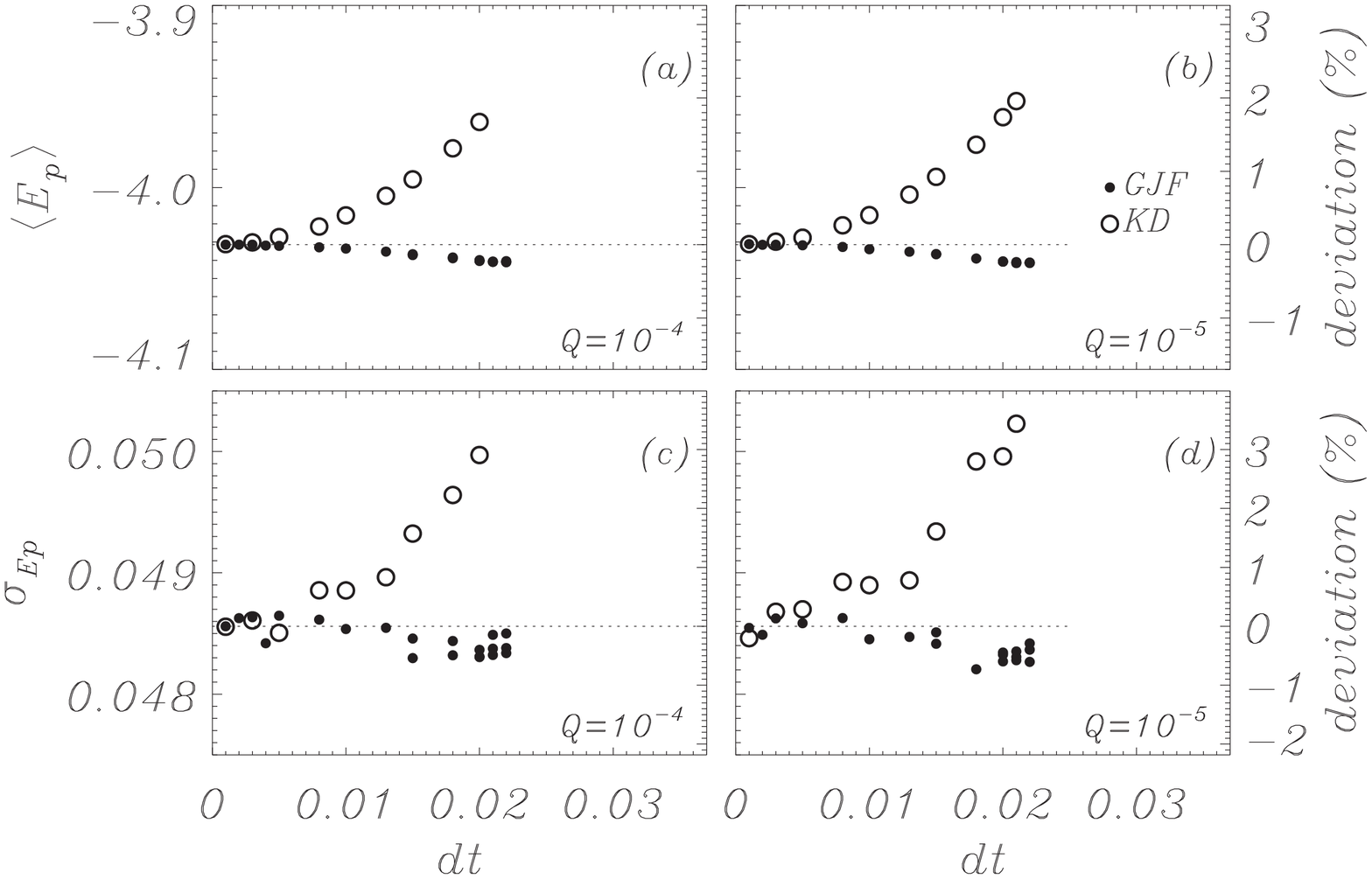}}
\end{flushleft}
\vspace{-0.5cm}
\caption{For $k_BT/E_0=0.7$ (liquid phase): Simulated average
potential energy $\langle{E_p}\rangle$ [(a) and (b)] and standard
deviation $\sigma_{E_p}$ [(c) and (d)] for $Q=10^{-4}$ [(a) and (c)]
and $Q=10^{-5}$ [(b) and (d)]. Markers represent the G-JF method of
this paper (solid $\bullet$) and the KD method (open
$\circ$). Horizontal dotted lines are leveled at $\langle{E_p}\rangle$
for $Q=10^{-4}$ and $dt=0.001$ [(a) and (b)], and at $\sigma_{E_p}$
for $Q=10^{-4}$ and $dt=0.001$ [(c) and (d)]. All figures show axes
with absolute quantities on the left and percentage deviation on the
right axes.}
\label{fig:fig_3pot}
\end{figure}

The $k_BT/E_0=0.7$ data for the potential energy are shown in
Fig.~\ref{fig:fig_3pot}. The KD results for this liquid phase exhibit
the typical BBK behavior that was also seen in Fig.~\ref{fig:fig_2pot}
for the solid phase at $k_BT/E_0=0.3$. In comparison, the potential
energy shows only a slight decrease in both average and fluctuations
for the G-JF method. It is again clear that the G-JF method produces
simulated matter with configurational properties nearly independent of
$dt$. The uncertainty on the acquired averages can be assessed from
the associated standard deviations and the averaging time. We here
also include a multiple of simulation data for the same parameters in
order to indicate the magnitude of the statistical error that should
be associated with the presented standard deviations.

\begin{figure}[t]
\begin{flushleft}
\scalebox{0.37}{\centering \includegraphics{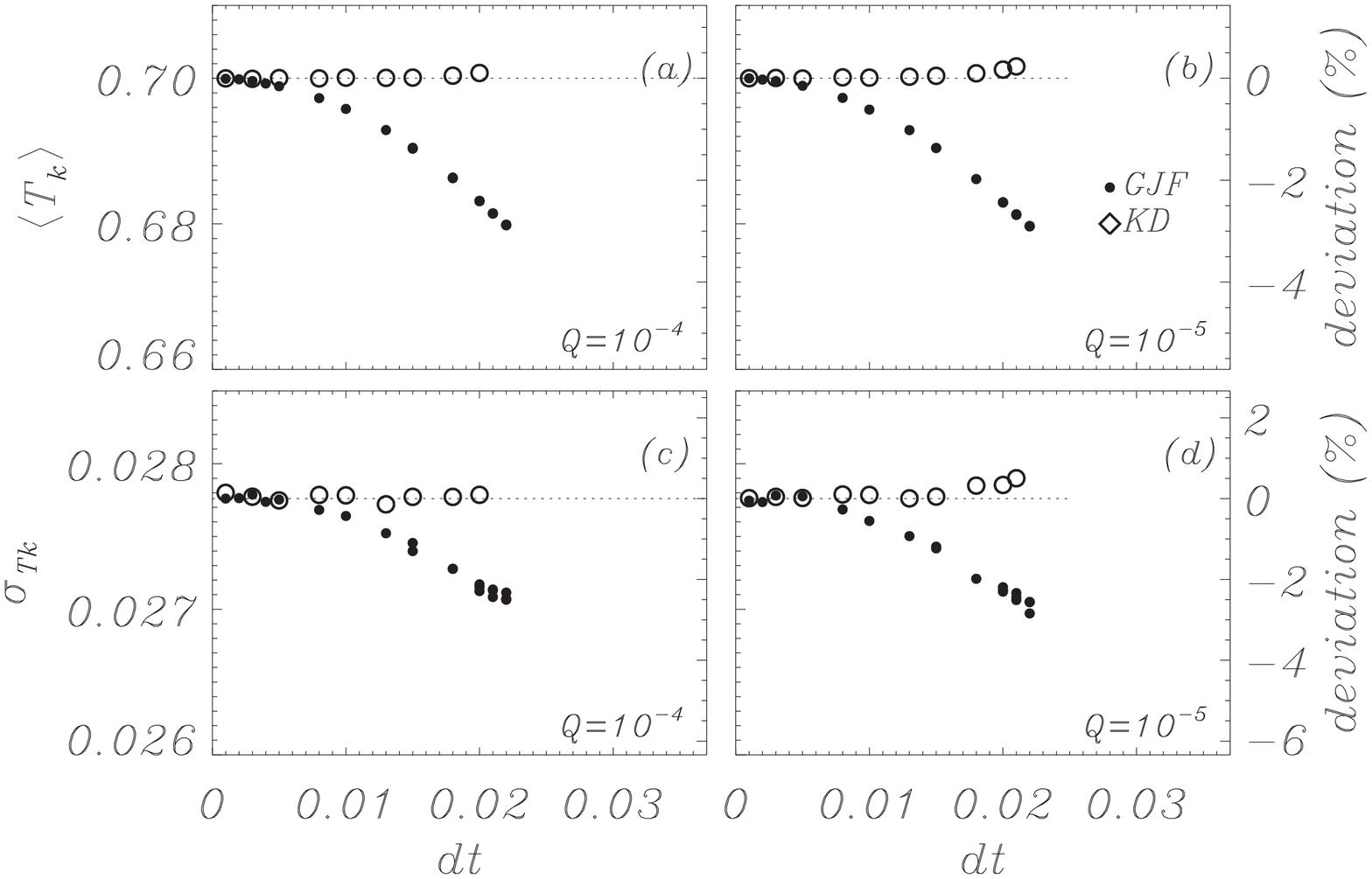}}
\end{flushleft}
\vspace{-0.5cm}
\caption{For $k_BT/E_0=0.7$ (liquid phase): Simulated average kinetic
temperature $\langle{T_k}\rangle$ [(a) and (b)] (from
Eq.~(\ref{eq:Eq_Tk})) and standard deviation $\sigma_{T_k}$ [(c) and
(d)] for $Q=10^{-4}$ [(a) and (c)] and $Q=10^{-5}$ [(b) and
(d)]. Markers represent the G-JF method of this paper (solid
$\bullet$) and the KD method (open $\circ$). Horizontal dotted lines
are leveled at $\langle{T_k}\rangle=0.3$ [(a) and (b)], and at
$\sigma_{T_k}$ for $Q=10^{-4}$ and $dt=0.001$ [(c) and (d)]. All
figures show axes with absolute quantities on the left and percentage
deviation on the right axes.}
\label{fig:fig_3kin}
\end{figure}

We confirm that the kinetic measurements of temperature and its
fluctuations behave similarly in the liquid and solid phases by
comparing Figs.~\ref{fig:fig_2kin} and \ref{fig:fig_3kin}. The latter
shows the data for $k_BT/E_0=0.7$, which again demonstrates the
signature of the momentum $mv^n$ not exactly being the conjugate
variable to $r^n$.  Thus, also for the liquid phase, we observe that
the calculated discrete-time kinetic temperature is progressively
short of the actual temperature of the configurational sampling
statistics that can be inferred from the potential energy measurements
in Figure \ref{fig:fig_3pot}.

\begin{figure}[t]
\begin{flushleft}
\scalebox{0.37}{\centering \includegraphics{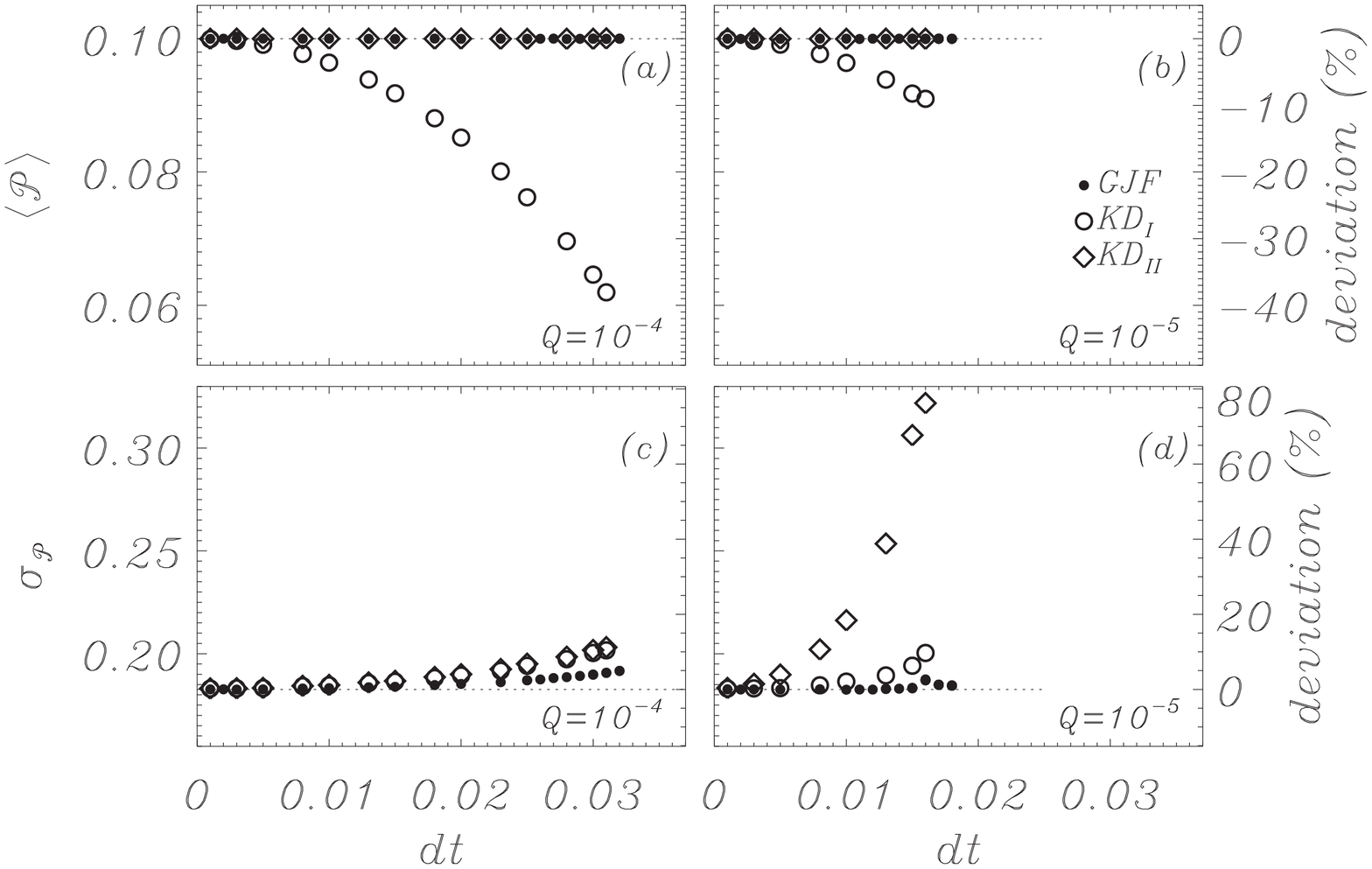}}
\end{flushleft}
\vspace{-0.5cm}
\caption{For $k_BT/E_0=0.3$ (solid phase): Simulated average pressure
$\langle{\cal P}\rangle$ [(a) and (b)] and standard deviation
$\sigma_{\cal P}$ [(c) and (d)] for $Q=10^{-4}$ [(a) and (c)] and
$Q=10^{-5}$ [(b) and (d)]. Markers represent the G-JF method of this
paper (solid $\bullet$), KD method using the instantaneous pressure from
Eq.~(\ref{eq:Eq_Inst_Press}) (open $\circ$), and KD method using
Eq.~(\ref{eq:Eq_KD_Press}) (open $\diamond$). Horizontal dotted lines are
leveled at $P=0.1$ [(a) and (b)], and at $\sigma_{\cal P}$ for
$Q=10^{-4}$ and $dt=0.001$ [(c) and (d)]. All figures show axes with
absolute quantities on the left and percentage deviation on the right
axes.}
\label{fig:fig_2P}
\end{figure}

We finally turn to investigating the pressure. Clearly, one should expect that the average
internal pressure $\langle{\cal P}\rangle$ is controlled and equal to
the imposed external pressure $P$, since this is the principal purpose
of the barostat.  
It is important to note that the internal pressure is defined
differently in the G-JF and KD methods (see discussion above in
section \ref{sec:method}). The former uses the canonical ensemble
definition [see Eq.~(\ref{eq:presgjf})]
\begin{eqnarray}
{\cal P}_{can}&=&\frac{1}{3V}\sum_{i=1}^N\bar{f}_i\cdot\bar{r}_i+
\frac{Nk_BT}{V},
\end{eqnarray}
while the latter targets the microcanonical ensemble expression [see Eq.~(\ref{eq:presandersen})]
\begin{eqnarray}
{\cal P}_{micro}&=&\frac{1}{3V}\sum_{i=1}^N\bar{f}_i\cdot\bar{r}_i
+\frac{1}{3V}\sum_{i=1}^Nm\bar{v}_i^2.
\label{eq:cont_time_press}
\end{eqnarray}
For each method we inspect the statistics of the relevant internal
pressure. The results for the solid phase simulations at
$k_BT/E_0=0.3$ are displayed in Figure \ref{fig:fig_2P}. From the
data, it is obvious that the imposed pressure is correctly adopted by the
G-JF method presented here. The KD method,
however, displays a curious and perhaps significant deviation from the
expected.  The origin of these deviations is the time reversible
discretization used in the KD method (see sequential steps (1)-(7) in
Sec.~V of Ref.~\cite{kolb}), that applies a trapezoidal approximation
[$(r^nf^n+r^{n+1}f^{n+1})/2$] to the configurational pressure
contribution [first term on rhs of Eq.~(\ref{eq:cont_time_press})],
while using a mid-point approximation ($v^{n+\frac{1}{2}}$) to the
kinetic part [second term on rhs of
Eq.~(\ref{eq:cont_time_press})]. Therefore, the discrete-time
instantaneous pressure
\begin{eqnarray}
{\cal P}^n_I & = &
\frac{1}{3V}\sum_{i=1}^N\bar{f}_i^n\cdot\bar{r}_i^n+\frac{1}{3V}\sum_{i=1}^Nm\left(\bar{v}_i^n\right)^2
, \label{eq:Eq_Inst_Press}
\end{eqnarray}
corresponding to the expression (\ref{eq:cont_time_press}) is
different from, and inconsistent with, the enforced pressure in the KD
method for $dt>0$. This inconsistency is visible in Figure
\ref{fig:fig_2P}, where the marker $\circ$ shows the average of the
instantaneous pressure calculated from
Eq.~(\ref{eq:Eq_Inst_Press}). The data exhibits a quadratically
increasing deviation between the enforced and measured internal
pressures as $dt$ is increased.  An measure of the internal pressure,
more consistent with the enforced value $P$, is found from
\begin{eqnarray}
{\cal P}^n_{II} & = &
\frac{1}{3V}\sum_{i=1}^N\bar{f}_i^n\cdot\bar{r}_i^n+\frac{1}{3V}\sum_{i=1}^Nm\left(\bar{v}_i^{n+\frac{1}{2}}\right)^2
, \label{eq:Eq_KD_Press}
\end{eqnarray}
which is shown by the markers $\diamond$. This measure of pressure
seems properly enforced for all time steps $dt$.

The fluctuations, defined as the standard deviation $\sigma_{\cal P}$,
of the internal pressure show very reasonable robustness of the G-JF
method against $dt$ variations, although we do observe up to about 5\%
error for $dt$ very close to the stability limit. In comparison, the
KD method shows larger deviations, especially for the measure of the
internal pressure defined by Eq.~(\ref{eq:Eq_KD_Press}). We thus
conclude that within the KD method, neither ${\cal P}^n_{I}$ nor
${\cal P}^n_{II}$ exhibit statistics that is insensitive to variations
in $dt$.

\begin{figure}[t]
\begin{flushleft}
\scalebox{0.37}{\centering \includegraphics{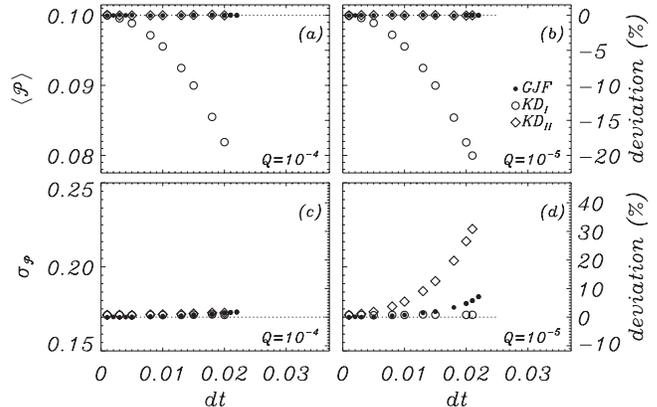}}
\end{flushleft}
\vspace{-0.5cm}
\caption{For $k_BT/E_0=0.7$ (liquid phase): Simulated average pressure
$\langle{\cal P}\rangle$ [(a) and (b)] and standard deviation
$\sigma_{\cal P}$ [(c) and (d)] for $Q=10^{-4}$ [(a) and (c)] and
$Q=10^{-5}$ [(b) and (d)]. Markers represent the G-JF method of this
paper (solid $\bullet$), KD method using the instantaneous pressure from
Eq.~(\ref{eq:Eq_Inst_Press}) (open $\circ$), and KD method using
Eq.~(\ref{eq:Eq_KD_Press}) (open $\diamond$). Horizontal dotted lines are
leveled at $P=0.1$ [(a) and (b)], and at $\sigma_{\cal P}$ for
$Q=10^{-4}$ and $dt=0.001$ [(c) and (d)]. All figures show axes with
absolute quantities on the left and percentage deviation on the right
axes.}
\label{fig:fig_3P}
\end{figure}

Figure \ref{fig:fig_3P} shows the acquired statistics of the measured
pressure and its fluctuations as a function of the time step for the
liquid phase at $k_BT/E_0=0.7$. The
overall behavior of the methods is the same for liquid and solid
phases with direct averages of the instantaneous pressure
Eq.~(\ref{eq:Eq_Inst_Press}) being significantly depressed for the KD
method as $dt$ is increased. We also see that the fluctuations of the
KD pressure is fairly independent of $dt$ for $Q=10^{-4}$, while the fluctuations of
the pressure ${\cal P}^n_{II}$ increase dramatically for
$Q=10^{-5}$. The G-JF method is generally robust, although we do
observe some increase in pressure fluctuations for $Q=10^{-5}$.

We note that improved statistical accuracy of kinetics using half-step velocities
$v^{n+\frac{1}{2}}$ in the so-called leap-frog versions of the Verlet
method have been investigated \cite{holian95,hoover13} for
deterministic Nos{\'e}-Hoover control of temperature and
pressure. However, while the half-step velocities may be able to
produce better consistency for the averaged kinetic temperature in
deterministic dynamics, these approaches may neither translate to
stochastic dynamics nor resolve the fundamental question of
calculating instantaneous pressure (at times $t_n$) for statistical averages
and fluctuations, as illustrated above.

\section{Discussion}
\label{sec:summary}
We have presented and demonstrated a new thermostat-barostat pair for
simulating atomic and molecular dynamics with periodic boundary
conditions.  The new G-JF method is simple and stable, and simulations
of thermodynamic properties produce data with very little dependency
on the applied numerical time step. We have investigated the method in
the context of two characteristic models - a one-dimensional toy model
with known statistical solutions, and the classical three-dimensional
Lennard-Jones material, simulated in both crystalline and liquid
phases.  In all cases the G-JF method behaves extremely well for
measured averages as well as for their fluctuations. In comparison,
the state-of-the-art KD method, which is also representative of other
commonly used methods, may exhibit significant deviations in both
averages and fluctuations for increasing time steps. As we have
emphasized throughout this paper, and specifically in Appendix
\ref{appendix_b}, it is crucial to appreciate that discrete time
invalidates the conjugate relationship between the coordinate $r$ and
its simulated velocity $v$. Consequently, one {\it cannot} expect
accurate simulation measures for both configurational and kinetic
quantities using any given method. This interesting and essential
feature becomes apparent when comparing the behavior of kinetic and
potential energies as a function of time step variations. 
We submit that the G-JF thermostat and barostat are advantageous in
that they consistently provide proper configurational properties (such
as Boltzmann distributions, Einstein diffusion, potential energy,
pressure, system volume, as well as their fluctuations), while leaving
kinetic measures (such as measured kinetic energy and the derived
kinetic temperature) with predictable deviations. In contrast, most
other methods (e.g., KD, Nos\'{e}-Hoover, etc.) enforce the expected
kinetic measures, thereby sacrificing the accuracy of proper
configurational sampling. The latter is unfortunate, since {\it most
molecular simulations are conducted in order to obtain configurational
information}.

We close by noting that the G-JF method is easily extended to
non-isotropic volume adjustments, and that the algorithm is not only
simple, but also in a form that makes it easy to implement into
existing molecular dynamics codes that have thermodynamic temperature
and pressure control. Specifically, the method can also conveniently
be expressed in the so-called leap-frog and position-Verlet forms, as
outlined for the thermostat in Ref.~\cite{gjf2}.

\acknowledgments This work was supported by the US Department of
Energy, project \# DE-NE0000536000, and by the Israel Science
Foundation (ISF), grant no.~1087/13.

\appendix
\section{Discrete-time relationship between position and velocity}
\label{appendix_b}
The continuous-time expectation, that the momentum $p^n=mv^n$ is the
conjugate variable to the spatial coordinate $r^n$, is {\it not}
fulfilled in discrete-time Verlet methods. This unfortunate
consequence of time discretization has significant implications for
the use and interpretation of simulations, and it can be illuminated
by considering a the simple analysis of a simulated harmonic oscillator \cite{hoover87},
\begin{eqnarray}
m\ddot{r} & = & -\kappa r \label{eq:Eq_harmonic_cont},
\end{eqnarray}
where $\kappa>0$ is a Hooke's spring constant. The continuous-time
 solution to this equation is, of course, $r(t)\propto\exp(\pm
 i\Omega_0 t)$ and $v(t)=\dot{r}=\pm i\Omega_0r(t)$, where
 $\Omega_0^2=\kappa/m$. (We use $i$ for complex notation in this
 Appendix.)
 
The discrete-time Verlet equations for Eq.~(\ref{eq:Eq_harmonic_cont}) are found from Eqs.~(\ref{eq:verlet_r}) and (\ref{eq:verlet_v}):
\begin{eqnarray}
 r^{n+1} & = & 2r^n-r^{n-1}-\frac{\kappa dt^2}{m} \, r^n\label{eq:Eq_harmonic_r}\\
 v^n & = & \frac{r^{n+1}-r^{n-1}}{2dt}\label{eq:Eq_harmonic_v}.
 \end{eqnarray}
The solution is
\begin{eqnarray}
r^n & \propto & \exp(\pm i\Omega_V\,n\,dt)\label{eq:Eq_harm_sol_r}
\end{eqnarray}
with the oscillation frequency $\Omega_V$ of the discrete-time Verlet
oscillator given by
\begin{eqnarray}
\sin(\Omega_Vdt) & = & \Omega_0dt\sqrt{1-\left(\frac{\Omega_0dt}{2}\right)^2}\label{eq:Eq_sin_OV}\\
\cos(\Omega_Vdt) & = & 1-\frac{\Omega_0^2dt^2}{2}\label{eq:Eq_cos_OV}
\end{eqnarray}
for $\Omega_0dt\le2$ (see, e.g., Refs.~\cite{gjf,hoover87}).  Inserting
Eq.~(\ref{eq:Eq_harm_sol_r}) into Eq.~(\ref{eq:Eq_harmonic_v}), we get
the Verlet velocity
\begin{eqnarray}
v^n & = & \pm\frac{\sin(\Omega_Vdt)}{\Omega_Vdt} \, i \Omega_V \, r^n\label{eq:Eq_harmonic_conj},
\end{eqnarray}
which, when compared to Eq.~(\ref{eq:Eq_harm_sol_r}), shows that the
Verlet velocity is {\it always} depressed in magnitude compared to the
true velocity of $r^n$ when $\Omega_Vdt\neq0$. For the extreme case of
the stability limit, $\Omega_0dt=2$ ($\Omega_Vdt = \pi$), we find
$v^n=0$, even if $r^n=(-1)^n r^0\neq0$ changes every time step. It is
obvious from the above analysis that the Verlet velocity is not an
accurate representation of the discrete time trajectory, and that any
kinetic measure, such as kinetic energy, temperature or diffusion derived from
velocity autocorrelations, cannot reliably
represent the behavior of the simulated configurational evolution (see also discussion
in the appendix of Ref.~\cite{holian95}).


\begin{thebibliography}{99}
\bibitem{velocity_diff} Notice the importance of using the Einstein
expression for diffusion, since this form directly measures the
displacement of the coordinates. Using velocity autocorrelation for
estimating diffusion is valid only for $dt\rightarrow0$ for the
reasons discussed in Appendix \ref{appendix_b} and commented on in Ref.~\cite{gjf2}.
\end{thebibliography}
\end{document}